\documentclass[a4paper,fleqn,usenatbib]{mnras}

\usepackage[T1]{fontenc}
\usepackage{ae,aecompl}
\usepackage[utf8]{inputenc}

\usepackage{graphicx}	% Including figure files
\usepackage{amsmath}	% Advanced maths commands
\usepackage{amssymb}	% Extra maths symbols
\usepackage{import}
\usepackage{lmodern}
\RequirePackage{fix-cm}

\usepackage{numprint}
\usepackage{SIunits}

\usepackage{grffile}
\newcommand{\rlight}{r_{\rm L}}

\newcommand{\rot}{\mathbf{\nabla} \times}

\DeclareRobustCommand{\rchi}{{\mathpalette\irchi\relax}}
\newcommand{\irchi}[2]{\raisebox{\depth}{$#1\chi$}} % inner command, used by \rchi

\title[Quadri-dipolar stars in vacuum]{Magnetic quadri-dipolar stars rotating in vacuum}

\author[J. P\'etri]{
J. P\'etri\thanks{E-mail: jerome.petri@astro.unistra.fr}
\\
Universit\'e de Strasbourg, CNRS, Observatoire astronomique de Strasbourg, UMR 7550, F-67000 Strasbourg, France.
}

\date{Accepted XXX. Received YYY; in original form ZZZ}

\pubyear{2019}

\begin{document}
\label{firstpage}
\pagerange{\pageref{firstpage}--\pageref{lastpage}}
\maketitle

\begin{abstract}
Main sequence stars and compact objects like white dwarfs and neutron stars are usually embedded in magnetic fields that strongly deviate from a pure dipole located right at the stellar centre. An off-centred dipole can sometimes better adjust existing data and offer a simple geometric picture to include multipolar fields. However, such configurations are usually to restrictive, limiting multipolar components to strength less than the underlying dipole. In this paper, we consider the most general lowest order multipolar combination given by a dipole and a quadrupole magnetic field association in vacuum. Following the general formalism for multipolar field computations, we derive the full electromagnetic field outside a rotating quadridipole. Exact analytical expressions for the Poynting flux and the electromagnetic kick are given. Such geometry is useful to study the magnetosphere of neutron stars for which more and more compelling observations reveals hints for at least quadridipolar fields. We also show that for sufficiently high quadrupole components at the stellar surface, the electromagnetic kick imprinted to a neutron star can reach thousands of km/s for a millisecond period at birth.
\end{abstract}

\begin{keywords}
	magnetic fields - methods: analytical - stars: neutron - stars: rotation - pulsars: general
\end{keywords}

%%%%%%%%%%%%%%%%%%%%%%%%%%%%%%%%%%%%%%%%%%%%%%%%%%

%%%%%%%%%%%%%%%%% BODY OF PAPER %%%%%%%%%%%%%%%%%%

\section{Introduction}

Stars, wherever from the main sequence or being compact objects, are usually rotating at various speeds and possess a magnetic field of various strengths, depending on their internal structure, reflecting the pressure and density conditions inside the star. The interplay between rotation and magnetic field induces an electric field that has far reaching consequences for the fate of massive stars. Magnetic fields indeed play a central role in the evolution of star and in their final fate into compact objects like white dwarfs, neutron stars and black holes. 

There exist increasing evidences that a star-centred magnetic dipole is unable to explain the wealth of observations in many types of stars. Sometimes an off-centred dipole suffices to reconcile the model with the data. For instance, for magnetic white dwarfs, radio polarization profiles are satisfactorily fitted by an off-centred dipole \citep{putney_off-centered_1995} or a dipole+quadrupole. For strongly magnetized neutron stars, the off-centring could explain the high velocity kick at birth \citep{harrison_acceleration_1975, lai_pulsar_2001}. The electromagnetic kick arises mainly from the interaction between a magnetic dipole and a magnetic quadrupole. This configuration can also explain the time delay between radio pulses and thermal X-rays \citep{petri_joint_2020}. For an off-centred dipole, the quadrupole strength is bounded by the distance~$d$ to the center of the star of radius~$R$. It is at most of the same magnitude as the dipole, linearly scaling with $\epsilon = d/R \lesssim1$. This explains why the kick cannot be as high as those observed in some neutron stars \citep{johnston_evidence_2005}. Therefore, to circumvent this limit, \cite{kojima_kick_2011} looked for less restrictive field topologies where the quadrupole is unconstrained and unrelated to the dipole. A general-relativistic description of a dipole+quadrupole system has been considered by \cite{gralla_inclined_2017}. Last but not least, the recent fitting of the polar cap thermal X-ray emission of the millisecond pulsar PSR~J0030+0451 confirms the view of a dipole+quadrupole field at the surface \citep{riley_nicer_2019, bilous_nicer_2019}. Force-free quadridipole magnetospheres are thus relevant to compute in order to extract radio and gamma-ray light-curves as tempted by \cite{chen_numerical_2020}. Plasma filled magnetospheres are the regime to tend to but in this exploratory work, we look for simple and analytical expressions to get more insight into the quadridipole topology for any geometry.

Different scenario are at hand to explain neutron star kicks like asymmetric neutrino emission during the collapse with possible spin-kick correlation, hydrodynamical instabilities leading to asymmetric explosion and electromagnetic thrust operating on a longer time scale. \cite{rankin_further_2007} further found evidence of spin-kick alignment by analysing polarization position angle variation. \cite{johnston_evidence_2007} also found a plausible alignment for 7 more pulsars.

In this paper we compute exact analytical expressions for the electromagnetic field in vacuum outside a rotating arbitrary quadridipole configuration. Sec.~\ref{sec:Modele} reminds the basic solutions for a single multipole of any order~$\ell$ and $m$. In Sec.~\ref{sec:Luminosite} the associated spin-down luminosity and electromagnetic force expected from the quadridipole are derived. We finish by a discussion on possible neutron star electromagnetic kicks at birth in light of observations in Sec.~\ref{sec:neutron_star_kick}. Conclusions are summarized in Sec.~\ref{sec:Conclusion}.

\section{Rotating magnetic multipoles}
\label{sec:Modele}

In this section, we remind the essential results found by \cite{petri_multipolar_2015} for single multipolar fields. We then specialize to the quadridipole configuration. The star is rotating at an angular speed~$\Omega$ and the transition region between the static zone and the wave zone occurs around the light-cylinder defined by $\rlight=c/\Omega$ meaning that inside $r\ll\rlight$ it resembles a static multipole in corotation and outside, at large distances $r\gg\rlight$ it corresponds to a plane electromagnetic wave radiating energy and angular momentum responsible for the spin down torque and kick.

\subsection{Analytical solutions}

The quadridipole model is built on the exact analytical expressions for an arbitrary rotating multipolar magnetic field in vacuum derived by \cite{petri_multipolar_2015}. He showed that the vacuum solution is fully determined by the radial magnetic field component at the stellar surface of radius~$R$. The vacuum electromagnetic field is expanded onto vector spherical harmonics $\mathbf{\Phi}_{\ell,m}$ and $\mathbf{\Psi}_{\ell,m}$ of order~$\ell$ and $m$, see \cite{petri_general-relativistic_2013} for definitions in a curved space-time. The magnetic part~$\mathbf{B}$ is represented by the constant coefficients~$a_{\ell,m}^{\rm B}$ whereas the electric part~$\mathbf{D}$ is represented by the constant coefficients~$a_{\ell,m}^{\rm D}$. An arbitrary solution of Maxwell equations in vacuum is therefore advantageously written in spherical polar coordinates~$(r,\vartheta,\varphi)$ in the form of outgoing waves as
\begin{subequations}
	\label{eq:Constantes}
	\begin{align}
	\label{eq:Solution_Generale_div_0_D}
	& \mathbf{D}(r,\vartheta,\varphi,t) = \sum_{\ell=1}^\infty \rot [a^{\rm D}_{\ell,0} \, \frac{\mathbf{\Phi}_{\ell,0}}{r^{\ell+1}}] + \\
	& \sum_{\ell=1}^\infty \sum_{|m|\leq-\ell}^{m\neq0}  \left( \rot [a^{\rm D}_{\ell,m} \, h_\ell^{(1)}(k_m\,r) \, \mathbf{\Phi}_{\ell,m} ] +
	 \nonumber \right. \\	 &  \left. 
	i \, \varepsilon_0 \, m \, \Omega \, a^{\rm B}_{\ell,m} \, h_\ell^{(1)}(k_m\,r) \, \mathbf{\Phi}_{\ell,m} \right) \, e^{-i\,m\,\Omega\,t} \nonumber \\
	\label{eq:Solution_Generale_div_0_B}
	& \mathbf{B}(r,\vartheta,\varphi,t) = \sum_{\ell=1}^\infty \rot [a^{\rm B}_{\ell,0} \, \frac{\mathbf{\Phi}_{\ell,0}}{r^{\ell+1}}] + \\
	& \sum_{\ell=1}^\infty\sum_{|m|\leq-\ell}^{m\neq0}  \left( \rot [a^{\rm B}_{\ell,m} \, h_\ell^{(1)}(k_m\,r) \, \mathbf{\Phi}_{\ell,m}] 
	\nonumber \right. \\ & \left. 
	- i \, \mu_0 \, m \, \Omega \, a^{\rm D}_{\ell,m} \, h_\ell^{(1)}(k_m\,r) \, \mathbf{\Phi}_{\ell,m} \right) \, e^{-i\,m\,\Omega\,t} \nonumber.
	\end{align}
\end{subequations}
The wave number is $k_m = m \, k = m/\rlight$. 
The constants $\{a^{\rm D}_{\ell,m}, a^{\rm B}_{\ell,m}\}$ depend on the boundary conditions imposed on the stellar surface. The divergencelessness is satisfied by construction because of the properties of the vector spherical harmonics $\mathbf{\Phi}_{\ell,m}$ and $\mathbf{\Psi}_{\ell,m}$. $h_\ell^{(1)}$ are the spherical Hankel functions imposing outgoing wave conditions \citep{arfken_mathematical_2005}. The constants of integration are given for an asymmetric mode $m>0$ by
\begin{subequations}
\begin{align}
\label{eq:aBlm}
a^{\rm B}_{\ell,m} & = \frac{f^{\rm B}_{\ell,m}(R)}{h_\ell^{(1)}(k_m\,R)}
\end{align}
and for an axisymmetric case~$m=0$ by
\begin{align}
\label{eq:aBl0}
a^{\rm B}_{\ell,0} = R^{\ell+1} \, f^{\rm B}_{\ell,0}(R)
\end{align}
where the static magnetic field is expanded into
\begin{equation}
\mathbf{B}_{\rm stat}(r,\vartheta,\varphi,t) = \sum_{\ell=1}^\infty\sum_{m=-\ell}^\ell \left( \rot [f^{\rm B}_{\ell,m}(r) \, \mathbf{\Phi}_{\ell,m} ] \, e^{-i\,m\,\Omega\,t} \right).
\end{equation}
Specializing to a single multipole, the two non-vanishing electric field coefficients $a^{\rm D}_{\ell,m}$ are for an asymmetric mode $m>0$
\label{eq:aDlm}
\begin{align}
a^{\rm D}_{\ell+1,m} \, \left.\partial_r ( r \, h_{\ell+1}^{(1)}(k_m\,r))\right|_{r=R} & = \\ 
\varepsilon_0 \, R \, \Omega \,  \sqrt{\ell\,(\ell+2)} \, J_{\ell+1,m} \, f^{\rm B}_{\ell,m}(R) \nonumber \\
a^{\rm D}_{\ell-1,m} \, \left.\partial_r ( r \, h_{\ell-1}^{(1)}(k_m\,r))\right|_{r=R} & = \\ 
- \varepsilon_0 \, R \, \Omega \, \sqrt{(\ell-1)\,(\ell+1)} \, J_{\ell,m} \, f^{\rm B}_{\ell,m}(R) \nonumber
\end{align}
where we introduced the numbers~$J_{\ell,m} = \sqrt{\frac{\ell^2-m^2}{4\,\ell^2-1}}$. Note that for $\ell=1$ only one solution exists. For the axisymmetric case $m=0$ we find
\begin{align}
\label{eq:aDl0}
	(\ell+1) \, a^{\rm D}_{\ell+1,0} & = - \varepsilon_0 \, R^{\ell+3} \, \Omega \,  \sqrt{\ell\,(\ell+2)} \, J_{\ell+1,0} \, f^{\rm B}_{\ell,0}(R) \\
	(\ell-1) \, a^{\rm D}_{\ell-1,0} & =   \varepsilon_0 \, R^{\ell+1} \, \Omega \, \sqrt{(\ell-1)\,(\ell+1)} \, J_{\ell,0} \, f^{\rm B}_{\ell,0}(R).
	\end{align}
\end{subequations}
These relations are found by imposing the right jump conditions at the surface, namely continuity of the radial magnetic field component and continuity of the tangential electric field component. From the set of single multipoles, we deduce all the relevant quantities like the spin down luminosity and the electromagnetic force exerted on the star assuming a perfect conductor inside as usually done for instance for neutron stars.

\subsection{Spindown and force}

\cite{petri_radiation_2016} has shown that the spin down luminosity~$L$ reduces to a simple sum of the asymmetric $m>0$ modes with constants of integration $\{a^{\rm D}_{\ell,m}, a^{\rm B}_{\ell,m}\}$ in such a way that
\begin{equation}
\label{eq:luminosite}
L = \frac{c}{2\,\mu_0} \, \sum_{\ell\geq1}^{m\neq0}  \left( |a^{\rm B}_{\ell,m}|^2 + \mu_0^2 \, c^2 \, |a^{\rm D}_{\ell,m}|^2 \right) .
\end{equation}
For a single multipole of order~$\ell$, most textbooks on radiation only take into account the magnetic part proportional to~$|a^{\rm B}_{\ell,m}|^2$ giving the simple scaling $L\propto \Omega^{2\ell+4}$. This however neglects the contribution from the electric part given by $\mu_0^2 \, c^2 \, |a^{\rm D}_{\ell,m}|^2$ which is significant for fast rotators close to the break up limit where centrifugal forces would blow up the star. For neutron stars with small period at birth, around one millisecond, these corrections are rather substantial, of the order 10\%, we keep therefore the exact expression Eq.~(\ref{eq:luminosite}) in our application to neutron star kicks in Sec.~\ref{sec:neutron_star_kick}. More importantly, the magnetic quadrupole will contribute to the same order of magnitude as the electric dipole, this explains why we retain such corrections to remain self-consistent. $L$ is a simple sum of decoupled single multipoles meaning that there is no interaction between multipoles of different order $\ell$ and $\ell' \neq \ell$ whatever $m$ and $m'$. In addition, fast rotation also implies a deformation of the stellar surface from a perfect sphere to an oblate spheroid. Such perturbations of the neutron star shape also contributes to variations in the spindown luminosity. Unfortunately, there exists no analytical solution for such a geometry. Nevertheless, a homogeneous, uniformly rotating Newtonian Maclaurin spheroid with eccentricity~$e$ \citep{chandrasekhar_ellipsoidal_1970} defined by the difference in polar radius~$R_p$ and equatorial radius~$R_e$
\begin{equation}\label{eq:excentricite}
e \approx \frac{R_e-R_p}{(R_e+r_p)/2}
\end{equation}
could serve as a simple approximation to guess the corrections to the Poynting flux. According to \cite{algendy_universality_2014}, the difference in equatorial and polar gravity amounts to about~20\%, reflecting in an eccentricity very similar to the one found by \cite{morsink_oblate_2007}. Eccentricities are about $e \approx 0.5$ for millisecond pulsars \citep{belvedere_uniformly_2014}. Using the spin down luminosity derived for instance by \cite{finn_spin-down_1990} assuming magnetic flux freezing within the oblate star, the correcting factor is
\begin{equation}%\label{key}
\frac{\upi^2}{4} \, \frac{(1-e^2)^{2/3}}{E(e)^2} \approx 1 + \frac{1}{2} \, e - \frac{37}{96} \, e^2 + o(e^2)
\end{equation}
where $E(e)$ is the complete elliptic integral of the second kind \citep{arfken_mathematical_2005}. Using estimates from \cite{algendy_universality_2014}, a good guess for the eccentricity is
\begin{equation}%\label{key}
e \approx \frac{R_e-R_p}{R_e} \approx 0.8 \, a^2 \, \frac{R_e\,c^2}{G\,M} .
\end{equation}
Therefore, the spindown corrections due to oblateness is of the order $e/2$ which is comparable to the electric quadrupole term, depending also on $a^2$, albeit a modulation due to the compactness. Consequently, regarding the Poynting flux, the deviation from a perfect sphere can be formally absorbed in the electric quadrupole radiation part, not impacting much the results of the paper.

The associated electromagnetic force is derived by another sum now involving interferences between multipoles of different order $\ell$ and $\ell' \neq \ell$ but with the same azimuthal number $m=m'$. After some algebra, the kick reduces to 
\begin{subequations}
\label{eq:force}
	\begin{align}
	F & = \frac{1}{2\,\mu_0} \, \sum_{\ell,\ell'\geq1}^{m\neq0} i^{\ell'-\ell} \, \left( ( a^{\rm B}_{\ell,m} \, a^{\rm{B}^*}_{\ell',m} + \mu_0^2 \, c^2 \, a^{\rm D}_{\ell,m} \, a^{\rm{D}^*}_{\ell',m} ) \, I_1 
	\right. \nonumber \\	
	+ & \left. \mu_0 \, c \, ( a^{\rm B}_{\ell,m} \, a^{\rm{D}^*}_{\ell',m} - a^{\rm{B}^*}_{\ell',m} \, a^{\rm D}_{\ell,m} ) \, I_3 \right)
	\end{align}
\end{subequations}
where the integrals~$I_1$ and~$I_3$ have been computed in \cite{petri_radiation_2016}, see also \cite{roberts_electromagnetic_1979}. They are given by
\begin{subequations}
	\begin{align}
	I_1 & = \frac{\sqrt{\ell\,(\ell+2)}}{\ell+1} \, J_{\ell+1,m} \, \delta_{\ell+1,\ell'} + \frac{\sqrt{(\ell-1)\,(\ell+1)}}{\ell} \, J_{\ell,m} \, \delta_{\ell-1,\ell'} \\
	I_3 & = \frac{i\,m}{\ell\,(\ell+1)} \, \delta_{\ell,\ell'}
	\end{align}
\end{subequations}
where $\delta_{ik}$ is the Kronecker symbol. It can be checked that the force $F$ is a real number involving only the imaginary part of products like $(a^{\rm B}_{\ell,m} \, a^{\rm{B}^*}_{\ell',m})$, $(a^{\rm D}_{\ell,m} \, a^{\rm{D}^*}_{\ell',m})$ and $(a^{\rm B}_{\ell,m} \, a^{\rm{D}^*}_{\ell',m})$ as already emphasized by \cite{roberts_electromagnetic_1979}. The terms involving the same multipoles $\ell=\ell'$ cancel as it should for single multipoles of order $(\ell,m)$. Due to the symmetry of these single multipoles, no force can be produced by such fields.

Consequently, we are led to the two key results of Eq.~(\ref{eq:luminosite}) and Eq.~(\ref{eq:force}) which are the most general expressions valid for any combination of multipoles in vacuum. From now on, we specialize to the dipole and quadrupole fields shortened as a quadridipole.

\subsection{Quadridipole field}

Taking the single multipole solutions from \cite{petri_multipolar_2015}, the dipole electromagnetic field is fully determined by the typical magnetic field strength $B_{\rm dip}$ and by the following constants
\begin{subequations}
	\label{eq:Deutsch}
	\begin{align}
	a_{1,0}^{\rm B} & = -\sqrt{\frac{8\,\upi}{3}} \, B_{\rm dip} \, R^3 \, \cos\rchi \\
	a_{1,1}^{\rm B} & = \sqrt{\frac{16\,\upi}{3}} \, \frac{B_{\rm dip} \, R}{h_1^{(1)}(k\,R)} \, \sin\rchi \\
	a_{2,0}^{\rm D} & = \sqrt{\frac{8\,\upi}{15}} \, \varepsilon_0 \, \Omega \, B_{\rm dip} \, R^5 \, \cos\rchi \\
	a_{2,1}^{\rm D} & = \sqrt{\frac{16\,\upi}{5}} \, \varepsilon_0 \, \frac{\Omega \, B_{\rm dip} \, R^2}{\left. \partial_r(r\,h_2^{(1)}(k\,r))\right|_{r=R}} \, \sin\rchi
	\end{align}
\end{subequations}
$\rchi\in[0,\upi]$ is the magnetic dipole inclination angle with respect to the rotation axis, taken to be along the $z$-axis.
The quadrupole field is determined by the typical magnetic field strength $B_{\rm quad}$ and by
\begin{subequations}
	\begin{align}
	a_{2,0}^{\rm B} & = q_{2,0} \, B_{\rm quad} \, R^4 \\
	a_{2,1}^{\rm B} & = q_{2,1} \, \frac{B_{\rm quad} \, R}{h_2^{(1)}(k_1\,R)} \\
	a_{2,2}^{\rm B} & = q_{2,2} \, \frac{B_{\rm quad} \, R}{h_2^{(1)}(k_2\,R)} \\
	a_{1,0}^{\rm D} & = \frac{2}{\sqrt{5}} \, \varepsilon_0 \, \Omega \, B_{\rm quad} \, R^4 \, q_{2,0} \\
	a_{1,1}^{\rm D} & = - \sqrt{\frac{3}{5}} \, \varepsilon_0 \, \frac{\Omega \, B_{\rm quad} \, R^2 \, q_{2,1}}{\left. \partial_r(r\,h_1^{(1)}(k_1\,r))\right|_{r=R}} \\
	a_{3,0}^{\rm D} & = -2\,\sqrt{\frac{2}{35}} \, \varepsilon_0 \, \Omega \, B_{\rm quad} \, R^6 \, q_{2,0} \\
	a_{3,1}^{\rm D} & = \frac{8}{\sqrt{35}} \, \varepsilon_0 \, \frac{\Omega \, B_{\rm quad} \, R^2 \, q_{2,1}}{\left. \partial_r(r\,h_3^{(1)}(k_1\,r))\right|_{r=R}} \\
	a_{3,2}^{\rm D} & = 2\,\sqrt{\frac{2}{7}} \, \varepsilon_0 \, \frac{\Omega \, B_{\rm quad} \, R^2 \, q_{2,2}}{\left. \partial_r(r\,h_3^{(1)}(k_2\,r))\right|_{r=R}} .
	\end{align}
\end{subequations}
The contribution of each quadrupolar component is weighted by the two angles~$\rchi_1, \rchi_2$ defined by 
\begin{subequations}
	\begin{align}
	q_{2,0} & = \sqrt{\frac{4\,\upi}{3}} \, \cos \rchi_1 \\
	q_{2,1} & = \sqrt{\frac{8\,\upi}{3}} \, \sin \rchi_1 \, \cos \rchi_2 \, e^{i\,\lambda_1} \\
	q_{2,2} & = \sqrt{\frac{8\,\upi}{3}} \, \sin \rchi_1 \, \sin \rchi_2 \, e^{2\,i\,\lambda_2}.
	\end{align}
\end{subequations}
where $\rchi_1\in[0,\upi]$ and $\rchi_2\in[0,2\,\upi]$. The angles $\lambda_1, \lambda_2$ fix the orientation of the $m=1$ and $m=2$ quadrupole with respect to a fiducial plane defined by the rotation axis and the magnetic dipole moment (assumed to be located in the $\varphi=0$ plane).

To summarize, all the field components are exactly known, the whole field topology being imposed by the seven parameters $(B_{\rm dip}, B_{\rm quad}, \rchi, \rchi_1, \rchi_2, \lambda_1, \lambda_2)$. $B_{\rm dip}$ sets the physical magnetic field strength not impacting the geometrical configuration therefore reducing the number of parameters for the geometry to six, five angles and the relative quadrupole strength $B_{\rm quad}/B_{\rm dip}$. We now apply these results to the secular evolution of a star rotating in vacuum.

\section{Star secular evolution}
\label{sec:Luminosite}

Some quantities are particularly relevant for the study of the stellar rotational evolution and electromagnetic kick. Although the exact analytical expression for the luminosity and force are readily found by straightforward application of the above formulas, we find it more pertinent to pick out meaningful limits depending on the geometry and rotation rate in order to emphasize the formal dependence on the six parameters.

\subsection{Spin down luminosity}

Let us start with the spin down rate, introducing the ratio between quadrupolar and dipolar field strength by $X=B_{\rm quad}/B_{\rm dip}$, the luminosity, to the lowest order in spin parameter $a=R/\rlight \lesssim 1$, can be compared to the perpendicular point dipole Poynting flux 
\begin{equation}%\label{key}
L_\perp = \frac{8\,\upi}{3\,\mu_0\,c^3} \, \Omega^4 \, B_{\rm dip}^2 \, R^6 
\end{equation}
such that
\begin{equation}%\label{key}
\frac{L}{L_\perp} = 
(1-a^2) \, \sin ^2 \rchi + %\\
 \frac{8}{45} \, a^2 \, X^2 \, \sin^2\rchi_1 \, \left(11-9 \, \cos 2 \rchi_2 \right) + o(a^2) .
%& a^4 \left(\frac{61 \sin^2 \rchi}{60}+\frac{1}{135} X^2 \sin^2 \rchi _1 \, \left(339 \cos \left(2 \rchi_2 \right) - 301 \right) \right) + O\left(a^5\right)
\end{equation}
Note that we use little-$o$ notations with a small~$o(x)$ not a big-$O$ notation with~$O(x)$, meaning that the expression is correct up to the order $a^2$, written as~$o(a^2)$ and not dominated by corrections starting at order~$a^2$ which would be written $O(a^2)$ (from Edmund Landau notations, a German mathematician).

The luminosity is independent of $\lambda_1,\lambda_2$ because it is unrelated to possible interaction between different multipoles and therefore insensitive to the relative orientation of each component. Single multipoles contribute separately to the total luminosity. The quadrupole contribution scales as $a^2\,X^2$ compared to a point dipole but for a finite size star, we need to retain also the $(1-a^2)$ correction of the dipole to be fully consistent. This term was neglected by \cite{kojima_kick_2011}. If the star is oblate due to rapid rotation close to break up, the deformation can be accounted for in the luminosity by adding another correcting factor scaling like~$a^2$, proportional to the compactness.

Note that the quadrupole luminosity dominates the electromagnetic braking only if the quadrupole strength in the wave zone dominates the magneto-dipole radiation in the regime where $a^2\,X^2 \gg 1$ and not when the quadrupole strength dominates at the surface $X^2 \gg 1$. The factor $a^2$ arises because the quadrupole field decrease much faster with radius than the dipole and the relevant field strength for radiation is located at the light-cylinder, not at the surface. This remark has profound implications for the stellar magnetic field estimates from the measured $P$ and $\dot{P}$ as discussed in depth by \cite{petri_illusion_2019}.

\subsection{Electromagnetic kick}
%\label{sec:Kick}

To get an electromagnetic force requires some coupling between several multipoles of different orders $\ell\neq\ell'$ in order to produce a kick. Again, to the lowest order in the spin parameter~$a$, it is expressed as
\begin{multline}%\label{key}
\frac{c\,F}{L_\perp} = \frac{1}{45} \, \sqrt{\frac{2}{15}} \, a \, X \, \sin \rchi \, \sin \rchi_1 \, \cos \rchi_2 \, \\
 \left( 75 \, a^3 \, \cos \lambda_1 + (90 - 7\, a^2) \, \sin \lambda_1  + o(a^3) \right).
\end{multline}
As expected it depends on the relative orientation between the quadrupole $(\ell,m)=(2,1)$ and the dipole $(\ell,m)=(1,1)$ through the angle~$\lambda_1$. Note however that a kick exists even for $\lambda_1=0$, but its formal dependence on $a$ is of higher order, $O(a^3)$ compared to $O(1)$ or $O(\Omega^3)$ instead of $O(1)$ (here we use big-$O$ notations). However, it is independent of the $(\ell,m)=(2,2)$ orientation because no term involving $\lambda_2$ appears. Comparing to the force experienced by an off-centred dipole \citep{petri_radiation_2016}, the off-centring $\epsilon\leq1$ has been replaced by $X$ which is unconstrained and possibly $X\gg1$. Consequently, an electromagnetic kick produced by a quadridipole becomes an interesting alternative to the supernova scenario to produce high spatial neutron star velocities for $X\gtrsim1$, a regime excluded for an off-centred dipole. Moreover this mechanism offers a natural explanation for the spin-velocity alignment observed in several systems \citep{rankin_further_2007,johnston_evidence_2007}.

In the last section, we show that such quadridipole configuration is indeed very effective in producing high kick velocities in neutron stars for millisecond periods at birth.

\section{Neutron star kick}
\label{sec:neutron_star_kick}

\subsection{A simple kick model}

The force and luminosity possess an intricate dependence on the unknown geometry of the system and on the initial rotation period and relative multipolar field strength symbolized by~$X$. In order to get a better insight, avoiding complications due to these uncertainties, we explore the impact of the spin parameter~$a$ and of the relative field strength~$X$ only, adjusting the geometry to the most favourable case. We therefore get an upper limit of the kick velocity. Let us take the particular geometry with $\rchi = \rchi_1 = \lambda_1 = \upi/2$ and $\rchi_2=0$. 
In this special case of an exact perpendicular magnetic dipole rotator, the torque exerted on the star does not align the magnetic axis towards the rotation axis because the time scale for this alignment becomes infinite \citep{michel_alignment_1970}. Therefore, we neglect this effect in the discussion. Let us however give an estimate of the alignment impact on the final kick velocity in the following lines.

For a general oblique dominant dipole configuration the rotation rate and inclination angle evolve according to
\begin{subequations}\label{eq:edo_alignment}
\begin{align}
I \, \Omega \, \frac{d\Omega}{dt} & = - L_\perp \, \sin^2\rchi \\
I \, \Omega^2 \, \frac{d\rchi}{dt} & = - L_\perp \, \cos\rchi \, \sin\rchi
\end{align}
\end{subequations}
where $I$ is the stellar moment of inertia. The joint evolution satisfies the constrain
\begin{equation}%\label{key}
\Omega \, \cos \rchi = \Omega_0 \, \cos \rchi_0
\end{equation}
where $\Omega_0$ and $\rchi_0$ are the initial rotation speed and inclination.
Introducing the time scale
\begin{equation}%\label{key}
\tau = \frac{I\,\Omega^4}{L_\perp \, \Omega_0^2 \, \cos \rchi_0^2}
\end{equation}
diverging for a perpendicular rotator ($\rchi_0=90\degr$) the solution for the angle and the rotation reads 
\begin{subequations}\label{eq:alignement}
\begin{align}
\sin\rchi & = e^{-t/\tau} \, \sin\rchi_0 \\
\Omega & = \Omega_0 \, \frac{\cos\rchi_0}{ \sqrt{ 1 - e^{-2\,t/\tau} \, \sin^2\rchi_0}} .
\end{align}
\end{subequations}
Thus, if the magnetic alignment is taking into account for an arbitrary oblique rotator in vacuum, the time scale for a significant change in the inclination angle is comparable to the spin down time scale. Therefore because the electromagnetic kick effect prevails in the era before the magnetic braking occurs, the alignment process only weakly decreases the kick velocity estimate. If the neutron star is surrounded by a plasma, like in the force-free model, the same qualitative conclusions apply.

\begin{figure}
	\centering
	\includegraphics[width=0.9\linewidth]{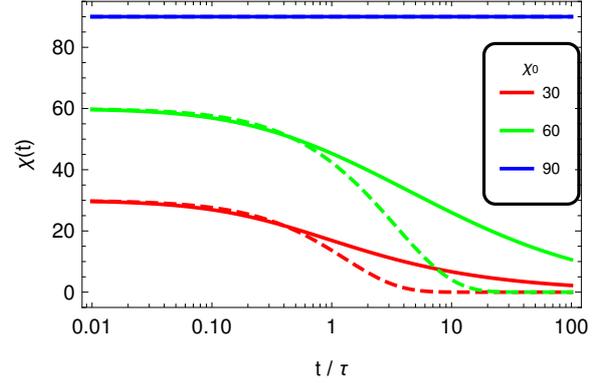}
	\caption{Evolution of the magnetic dipole inclination angle~$\rchi(t)$ with time for vacuum radiation, in dashed curves, and force-free radiation, in solid lines. The initial angle is $\rchi_0=\{30\degr,60\degr,90\degr\}$ respectively in red, green and blue.}
	\label{fig:angle_alignement}
\end{figure}

\begin{figure}
	\centering
	\includegraphics[width=0.9\linewidth]{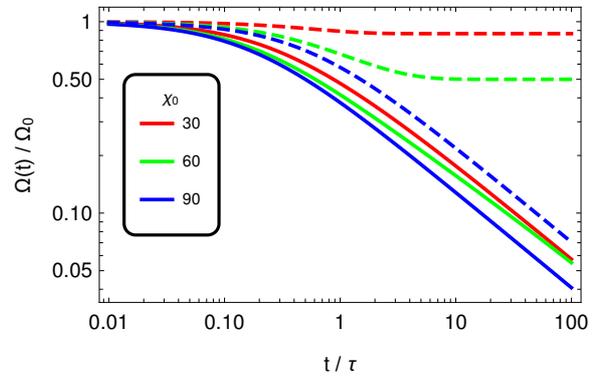}
	\caption{Evolution of the rotation rate~$\Omega(t)$ with time for vacuum radiation, in dashed curves, and force-free radiation, in solid lines. The initial angle is $\rchi_0=\{30\degr,60\degr,90\degr\}$ respectively in red, green and blue.}
	\label{fig:vitesse_alignement}
\end{figure}

Indeed, Fig.~\ref{fig:angle_alignement} shows the evolution of the magnetic dipole inclination angle with time for the vacuum (dashed lines) and force-free (solid lines) model. Significant decrease towards alignment start at the spindown timescale in both models but with a much slower decrease for the force-free case which goes as $\rchi(t) \propto t^{-1/2}$ contrary to an exponential vanishing as $\rchi(t) \propto e^{-t/\tau}$ for the vacuum case, Eq.~(\ref{eq:alignement}).
Fig.~\ref{fig:vitesse_alignement} shows the evolution of the rotation rate with time. The star rotation does not stop in the vacuum model if $\chi<90\degr$. Only the force-free model brakes the star until it tends to rest with $\Omega(t) \propto t^{-1/2}$ because an aligned rotator surrounded by a plasma also radiates. However, in all cases the typical time scale for magnetic braking to set in is again the spindown time scale~$\tau$. A quantitative analysis is proposed below after investigation of the general orthogonal case.

\begin{figure}
	\centering
	\includegraphics[width=0.9\linewidth]{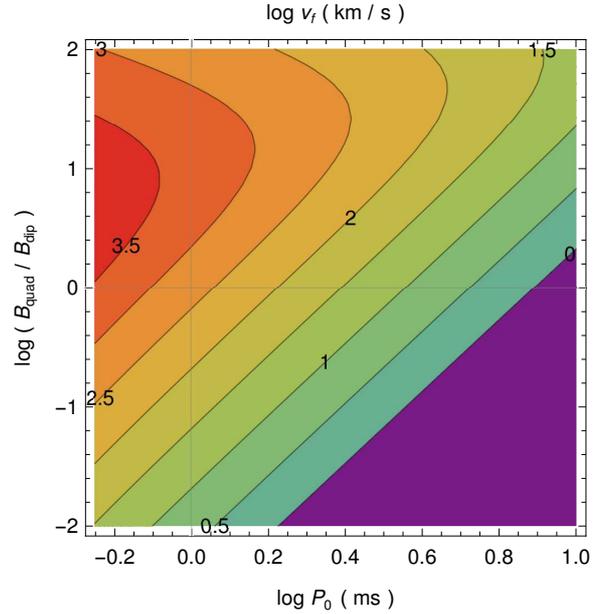}
	\caption{Isocontours of the final kick velocity expressed in logarithmic scale $\log v_f$ and in units of km/s, depending on the initial period~$P_0$ and relative quadrupole magnitude $X$.}
	\label{fig:vfinal}
\end{figure}

For an arbitrary rotation rate and field strengths with $\rchi = \rchi_1 = \lambda_1 = \upi/2$ and $\rchi_2=0$, the luminosity becomes 
\begin{equation}\label{eq:luminosite_max}
L = \left( 1 - a^2 + \frac{16}{45} \, X^2 \, a^2 \right) \, L_\perp 
\end{equation}
and the electromagnetic kick to lowest order neglecting small corrections in powers of~$a$ reads
\begin{equation}%\label{key}
F = 2 \, \sqrt{\frac{2}{15}} \, a \, X %\, \left( 1 - \frac{7}{90} \, a^2 \right) 
\, \frac{L_\perp}{c}.
\end{equation}
The time evolution of the kick velocity becomes
\begin{equation}\label{eq:vitesse}
v(t) = \int_{0}^{t} \frac{F(t)}{M} \, dt.%= 2 \, \sqrt{\frac{2}{15}} \, \frac{X}{M\,c} \, \int_{0}^{t} a \, \left( 1 - \frac{7}{90} \, a^2\right) \, L_{\perp}(t) \, dt .
\end{equation}
Changing the integration variable to $\Omega$ by noting that $d\Omega = \dot{\Omega} \, dt$ we have 
\begin{equation}%\label{key}
\frac{v(t)}{c} = 2 \, \sqrt{\frac{2}{15}} \, \frac{X}{M\,c^2} \, \int_{\Omega_0}^{\Omega} a \, %\left( 1 - \frac{7}{90} \, a^2\right) \,
 L_{\perp}(\Omega) \, \frac{d\Omega}{\dot\Omega} .
\end{equation}
Assuming a homogeneous neutron star with moment of inertia $I=\frac{2}{5} \, M \, R^2$, noting that $L = I\,\Omega\,\dot{\Omega}$, a primitive is
\begin{multline}\label{eq:betaAX}
\beta(a,X) = \frac{12 \, \sqrt{6} \, X}{5 \, \left(16 \, X^2-45 \right)^{3/2}} \\ \, \left[a \, \sqrt{80\, X^2-225} - 15 \, \arctan\left(\frac{1}{3} a \sqrt{\frac{16 \,X^2}{5}-9}\right)\right]
\end{multline}
and the time evolution for the kick
\begin{equation}%\label{key}
\frac{v(t)}{c} = \beta(a(t),X) - \beta(a_0,X)
\end{equation}
$a_0$ being the initial spin parameter at birth. The final kick velocity $v_f$ is reached whenever $a(t)=0$ therefore it becomes
\begin{equation}\label{eq:VFinal}
\beta_f = - \beta(a_0,X).
\end{equation}
Eq.~(\ref{eq:betaAX}) seems undefined whenever $X^2 \leq 45/16$ due to the square root in the $\arctan$ function and due to the vanishing denominator for $X^2 = 45/16$. However, for $X^2 < 45/16$, the square root in the numerator and denominator of the last term containing the $\arctan$ function becomes imaginary such that $\sqrt{X^2-45/16} = i\,x$ with $x\in \mathbb{R}$. Due to the relation between the inverse trigonometric and the inverse hyperbolic functions given by $\textrm{argth}(x)=i\,\arctan(-i\,x)$, Eq.~(\ref{eq:betaAX}) is well defined for $X^2 < 45/16$. Finally in the special case where $X^2 = 45/16$, the function~(\ref{eq:betaAX}) tends to 
\begin{equation}%\label{key}
\frac{1}{5} \, \sqrt{\frac{2}{3}} \, a_0^3 
\end{equation}
which is also well defined for any $a_0$. It can be checked that the function in Eq.~(\ref{eq:betaAX}) is continuous for all $a>0$ and all $x>0$.
It is shown in Fig.~\ref{fig:vfinal} as a contour plot depending on the initial period~$P_0$ and the quadrupole strength~$X$ on a logarithmic scale. For a 1~ms period at birth, the maximum allowed velocity is about 2000~km/s. In the limit of weak quadrupoles $X\ll1$, the kick velocity simplifies into
\begin{equation}%\label{key}
\beta_f =\frac{4}{5} \sqrt{\frac{2}{15}} \, X \, ( \textrm{argth}\,a_0 - a_0 ) 
\end{equation}
and for low initial rotational rate $a\ll1$ it further simplifies to
\begin{equation}\label{eq:vitesse_finale_AX_faible}
\beta_f = \frac{4}{15} \sqrt{\frac{2}{15}} \, a_0^3 \, X .
\end{equation}
In the extreme limit of dominant quadrupoles $X\gg1$, it reaches asymptotically
\begin{equation}%\label{key}
\frac{3}{2} \, \sqrt{\frac{3}{10}} \, \frac{a_0}{X}.
\end{equation}
Too much a strong quadrupole is counter-productive because the quadrupole radiation for $X\gg1$ dominates and considerably shortens the acceleration time (of the order the spindown time) which becomes vanishingly small.

In order to evaluate more quantitatively the impact of the alignment on the electromagnetic kick efficiency, let us compute the final speed in case of a small quadrupole component $X\ll1$ and a slow rotation rate $a\ll1$, extending the result of Eq.~(\ref{eq:vitesse_finale_AX_faible}) to a varying dipolar inclination angle. The electromagnetic kick is given approximately by
\begin{equation}%\label{key}
F = 2 \, \sqrt{\frac{2}{15}} \, a \, X %\, \left( 1 - \frac{7}{90} \, a^2 \right) 
\, \frac{L_\perp}{c} \, \sin\chi.
\end{equation}
Integrating Eq.~(\ref{eq:vitesse}) by a change of integration variable as
\begin{equation}
v(t) = \int_{0}^{t} \frac{F(t)}{M} \,  \frac{dt}{d\rchi}\,d\rchi
\end{equation}
according to the inclination angle evolution Eq.~(\ref{eq:edo_alignment}) we can perform the integration with respect to the initial obliquity~$\rchi_0$ down to alignment $\rchi=0$ finding
\begin{subequations}%\label{key}
\begin{align}
\beta_f & = - \frac{4}{5} \, \sqrt{\frac{2}{15}} \, a_0^3 \, X \, \cos^3\rchi_0 \, \int_{\rchi_0}^0 \frac{d\rchi}{\cos^4\rchi} \\
& = \frac{2}{15} \, \sqrt{\frac{2}{15}} \, a_0^3 \, X \, (3 \, \sin\rchi_0 + \sin 3\,\rchi_0 ).
\end{align}\end{subequations}
This expression is exactly Eq.~(\ref{eq:vitesse_finale_AX_faible}) when $\rchi_0=90\degr$. For neutron stars with initial inclinations far from an aligned rotator, the corrections are close to one. Therefore, the time evolution of the inclination angle remains negligible when computing the final kick velocity.  In all cases, the initial spin~$a_0$ and the quadrupole strength~$X$ fully determine the final kick velocity.

Can we set some upper limits for the initial period? For fast rotating neutron stars at birth, close to the break up limit we have according to general-relativistic corrections explained by \cite{friedman_implications_1989} and by \cite{haensel_equation_1995}
\begin{equation}%\label{key}
\Omega_{\rm max} \approx \numprint{0.67} \, \Omega_K
\end{equation}
corresponding to a fraction of the Keplerian frequency in Newtonian gravity~$\Omega_K = \sqrt{\frac{G\,M}{R^3}}$ and a spin parameter for a typical neutron star
\begin{equation}%\label{key}
a_{\rm max} \approx \numprint{0.28} \, \left( \frac{M}{1.4~M_\odot} \right)^{1/2} \, \left( \frac{R}{\numprint{12}~\SIunits{\kilo\meter}}\right)^{-1/2} .
\end{equation}
The maximum period before mass shedding is therefore around 1~\SIunits{\milli\second}. Corrections to the point dipole luminosity must be taken into account because $a\lesssim1$ and already for the dipole, corrections involve factors like $(1-a^2$). 

The maximum kick efficiency depends on $X$ for a given~$a$. Inspecting Fig.~\ref{fig:vfinal}, too a large quadrupole will decrease the maximum kick as well as too a low quadrupole. We can expect fast kicks above 1000~km/s if the quadrupole is comparable in magnitude to the dipole $X \sim 1$ for a millisecond initial period.
Whatever the strength of the quadrupole, high velocities larger than 1000~km/s require fast spinning neutron stars at birth with periods shorter than 2~ms.

Solutions to the electromagnetic kick problem are not unique. There exist a continuum of couple~$(a_0,X)$ satisfying the constrain. Of these, we highlight two special configurations: one with the smallest quadrupole component~$X$ and one with the largest initial period~$a_0$. We call the former solution the minimal choice and the latter solution the optimal choice because it follows the crest of the plot in Fig.~\ref{fig:vfinal} but then requires relatively large quadrupole magnitudes $X\gtrsim10$ although such values are not unrealistic.

\subsection{Indeterminacy of initial periods}

Are millisecond periods at birth realistic? The scenario depicted by \cite{heger_presupernova_2000} estimates the initial neutron star period from pre-supernova collapse models to be around those values of 1~ms. However, if angular moment is transported by magnetic field, \cite{heger_presupernova_2005} found an increase by one order of magnitude of this period. \cite{ott_spin_2006} also found initial spin periods about 1~ms but in contradiction to what is expected from observations of young pulsars inferring $P_0\gtrsim10$~ms. Young pulsars with smaller periods, like PSR~J0537-6910, are however not excluded to start their life with initial period less than 10~ms depending on their magnetic braking index \citep{marshall_discovery_1998}. For millisecond pulsars, the situation is more clear as \cite{ferrario_birth_2007} showed that they are usually born with periods close to the currently observed values.

According to \cite{atoyan_radio_1999}, if the magnetic field is decreasing with time, it is possible that the initial period of the Crab pulsar was around 3-5~ms, thus 10~times smaller than its current value, and not 19~ms if a constant braking index of $n=2.5$ is assumed \citep{manchester_pulsars_1977}. We must admit that guessing the initial period of young pulsars is a difficult and risky task because of our poor knowledge about the mechanisms responsible for the stellar spin down. Any inference must be taking with care, whatever its conclusion. We believe that millisecond period at birth are not excluded and take it as a serious hypothesis for the present work.

To properly fix the idea, let us quantitatively show the indeterminacy of initial periods of young pulsars even if the true age is known for instance by the associated supernova remanent age. In the simple spin down picture with a braking index of $n$, the true or current age $t_*$ is related to the characteristic age
\begin{equation}%\label{key}
\tau_{*} = - \frac{\Omega_*}{(n-1)\,\dot{\Omega}_*}
\end{equation}
by
\begin{equation}%\label{key}
t_* = \tau_{*} \, \left[ 1 - \left(\frac{\Omega_0}{\Omega_*}\right)^{n-1} \right].
\end{equation}
Knowing $t_*$ and $\tau_{*}$ from current observations, we can determine the initial rotation rate~$\Omega_0$ if a magnetic breaking model is assumed with a constant in time braking index~$n$. This method leads to birth periods well above 10~ms for pulsar/supernova associations. For the Crab with a braking index of $n=2.5$ we find $P_0=19$~ms. However, if the braking features are not constant in time, due to magnetic field decay, spin alignment or change in the moment of inertia among others, the guessed birth period can be much smaller. This was the idea of \cite{atoyan_radio_1999}.

\begin{figure}
	\centering
	\includegraphics[width=0.9\linewidth]{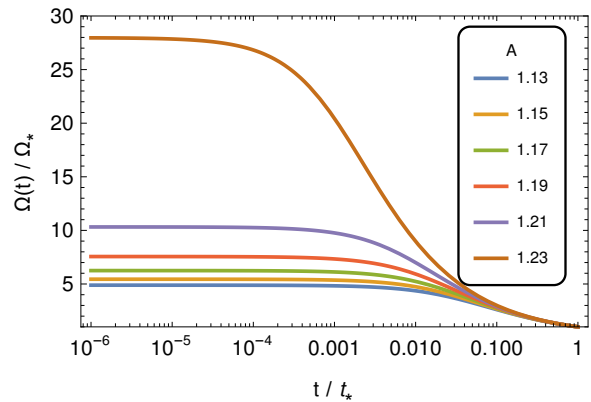}
	\caption{Evolution of the neutron star period $\Omega(t)$ compared to its current value~$\Omega_*$ at time $t_*=\tau_*/4$ for several values of the parameter~$A$ as in the legend and with a characteristic time $T=\tau_*/2$.}
	\label{fig:periode_initiale}
\end{figure}

Let us exemplify the general trend with a model taking into account the magnetic axis alignment for a force-free magnetosphere with $\rchi$ decreasing like $t^{-1/2}$ according to the time-dependent braking
\begin{equation}\label{eq:FreinageGeneral}
\dot{\Omega} = - K \, \left( 1 + \frac{A}{1+t/T} \right) \, \Omega^3
\end{equation}
mimicking a decrease in the obliquity for a force-free magnetosphere with characteristic time scale~$T$. Introducing the current typical time scale
\begin{equation}%\label{key}
\tau_* = \frac{1}{K\,\Omega_*^2} = - \frac{\Omega_*}{\left( 1 + \frac{A}{1+t_*/T} \right) \, \dot{\Omega}_*}
\end{equation}
and the variable~$\eta(t)=\Omega(t)/\Omega_*$, an exact solution with $\eta(t_*)=1$ is given by
\begin{equation}\label{eq:solution}
\eta(t) = \frac{1}{\sqrt{1+2\,(t-t_*)/\tau_* + 2\,A\,T/\tau_* \,\log\left(\frac{t+T}{t_*+T}\right)}}.
\end{equation}
The neutron star rotation rate at birth is therefore
\begin{equation}\label{eq:periode_initiale}
\Omega_0 = \frac{\Omega_*}{\sqrt{1-2\,t_*/\tau_* + 2\,A\,T/\tau_* \,\log\left(\frac{T}{t_*+T}\right)}}.
\end{equation}
For some combination of the free parameters $(A,T)$, both being positive, this ratio becomes very large $\Omega_0 \gg \Omega_*$. For instance, for a time scale $T$ associated to the spin down time scale $\tau_{*}$ because of magnetic alignment, we set $T=\tau_{*}/2$ and a true age of $t_* = \tau_{*}/4$, values of $A$ around $1.23$ lead to initial periods 30 times less than the current period, see Fig.~\ref{fig:periode_initiale}. Actually the birth period becomes very sensitive to this parameter around $A\approx1.23315$ where it tends to zero and the rotation rate diverges, $\Omega_0 \to + \infty$ whereas the constant coefficient case with $A=0$ leads only to an increase by a factor $\Omega_0/\Omega_* = \sqrt{2}$.

\subsection{Confrontation to several pulsar populations}

We confront our expectations against several populations of neutron stars. In a first subset, we consider neutron stars for which the three-dimensional space velocity can be inferred as demonstrated by \cite{cordes_neutron_1998}. In a second and third subset, we consider respectively young and millisecond pulsars extracted from the Australia Telescope National Facility (ATNF) catalogue with known $P$, $\dot{P}$ and proper motion without knowledge about the speed along the line of sight.

\subsubsection{Pulsars with inferred 3D velocity}

\begin{figure}
	\centering
	\includegraphics[width=0.9\linewidth]{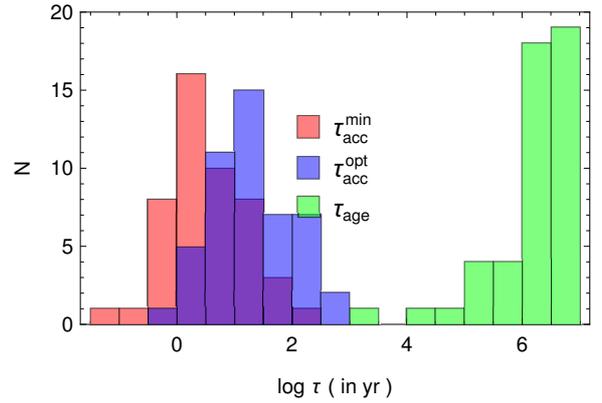}
	\caption{Acceleration time~$\tau_{\rm acc}^{\rm min}$ and~$\tau_{\rm acc}^{\rm opt}$ to the asymptotic kick velocity compared to the characteristic age.}
	\label{fig:temps_acceleration}
\end{figure}

We have selected the young pulsars studied in \cite{cordes_neutron_1998} because of the 3D velocity inference they gave. We found that 48 of these pulsars have known $P$ and $\dot{P}$ reported in the ATNF catalogue \citep{manchester_australia_2005}. They all possess a dipole magnetic field strength at the equator around \numprint{e8}-\numprint{e9}~\SIunits{\tesla} according to the magneto-dipole losses formula
\begin{equation}%\label{key}
B(\textrm{in T}) \approx \numprint{3.2e15} \, \sqrt{P(\textrm{in s})\,\dot{P}}
\end{equation}
where the period is measured in seconds and the field strength in T. Their characteristic age, assuming a single dipole model and values $(P,\dot{P})$ measured at present time, and defined by 
\begin{equation}%\label{key}
\tau_{\rm age} = \frac{P}{2\,\dot{P}}
\end{equation}
is also shown in Fig.~\ref{fig:temps_acceleration}. Most of them are older that millions of years expect for several of them being only aged about several thousands of years.

\begin{figure}
	\centering
	\includegraphics[width=0.9\linewidth]{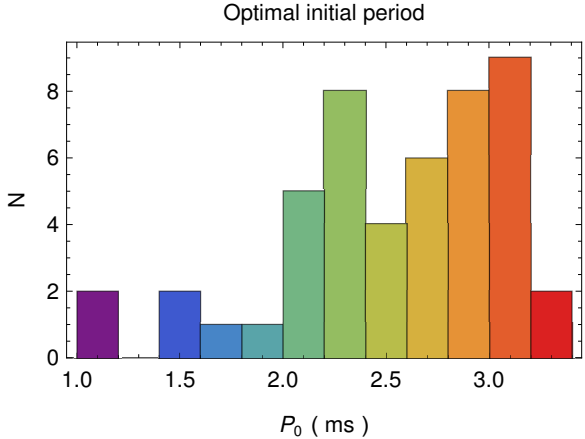}
	\caption{Expected distribution of initial neutron star spin periods~$P_0$ for the sample of 48~pulsars.}
	\label{fig:P_init_optimal}
\end{figure}

\begin{figure}
	\centering
	\includegraphics[width=0.9\linewidth]{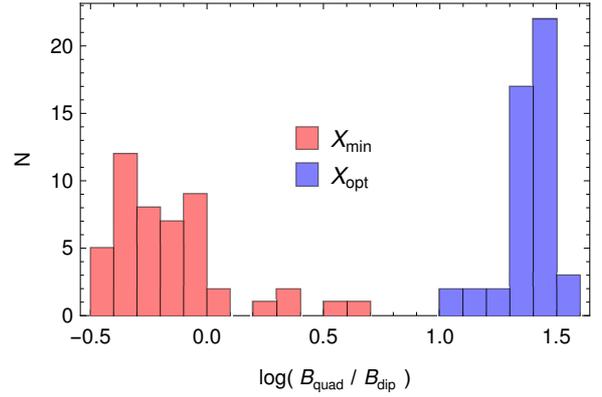}
	\caption{Optimal and minimum quadrupole strength~$X$ distribution of the 48~pulsars to explain the kick velocity for a period at birth, respectively shown in Fig.~\ref{fig:P_init_optimal} and for $P_0=1$~ms.}
	\label{fig:X_minimum}
\end{figure}

The histogram in Fig.~\ref{fig:P_init_optimal} show the expected distribution of largest possible initial periods~$P_0$ ranging from 1 to 3.5~ms for the sample of 48~pulsars assuming that their space velocity is known from the work of \cite{cordes_neutron_1998}. These values must be taken with caution because the radial component is basically unknown and the proper motion still rather uncertain. At least, we get a window for the initial period.
The counterpart of these largest period is to produce the largest quadrupole contributions as shown in blue bars in the histogram of Fig.~\ref{fig:X_minimum} where $X$ ranges from 10 to 35 times the dipole.

\begin{figure}
	\centering
	\includegraphics[width=0.9\linewidth]{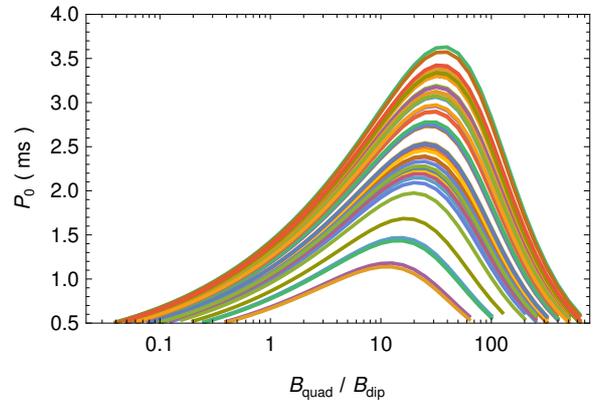}
	\caption{Initial spin period~$P_0$ depending on the quadrupole strength for the sample of 48~pulsars. Each curve with different colour depicts another pulsar.}
	\label{fig:P_init_sample}
\end{figure}

However, the solutions are degenerate because there is an infinite set of couples $(P_0,X)$ satisfying the imposed kick velocity. This is clearly seen in Fig.~\ref{fig:P_init_sample} where the relation between period and quadrupole relative magnitude is plotted for each pulsar in a different colour. What we called optimal choice corresponds to the solution found by the maximum on each curve. Nevertheless, the magnitude of the quadrupole can be decreased at the expense of increasing the spin of the neutron star.
For instance, assuming that all neutron stars are born with a period $P_0=1$~ms, the quadrupole distribution resembles the red bars in the histogram of Fig.~\ref{fig:X_minimum}. For most of them, the quadrupole is weaker then the dipole, in all cases it never exceeds 6 times the dipole. As a conclusion, even a modest quadrupole can efficiently kick up a neutron star during its early phase of spin down.

We finish by a discussion on the timescale necessary to achieve the final velocity. The characteristic time scale of the spin down depends on the spin down ratio at birth, according to Eq.~(\ref{eq:luminosite_max})
\begin{equation}%\label{key}
 \xi_0 = \frac{L_{\rm quad}}{L_{\rm dip}} \approx \frac{16}{45} \, a_0^2 \, X^2
\end{equation}
therefore not to be confused with $X$ which is only relating field strengths and not luminosities. Introducing the dipole and quadrupole characteristic time scales~$\tau_{\rm dip}$ and $\tau_{\rm quad} = \tau_{\rm dip}/\xi_0$ the total time scale becomes
\begin{equation}%\label{key}
\frac{1}{\tau_{\rm c}} = \frac{1}{\tau_{\rm dip}} +\frac{1}{\tau_{\rm quad}} .
\end{equation}
For non vanishing dipoles, we get
\begin{equation}%\label{key}
\tau_{\rm c} = -\frac{\Omega_0}{\dot{\Omega}_0} = \frac{\tau_{\rm dip}}{1+\xi_0} .
\end{equation}
The characteristic time scale is therefore diminishing for increasing quadrupole contribution. This scale can be compared to the traditional pulsar characteristic age given by $\tau_{\rm age}$. We choose to evaluate the ages from slow down rate at birth. If the quadrupole Poynting flux dominates $\xi_0\gg1$ the braking occurs on a much shorter time scale $\tau_{\rm c} \ll \tau_{\rm dip}$, and the electromagnetic kick becomes ineffective because of the too short duration of significant thrust phase.

\begin{figure}
	\centering
	\includegraphics[width=0.9\linewidth]{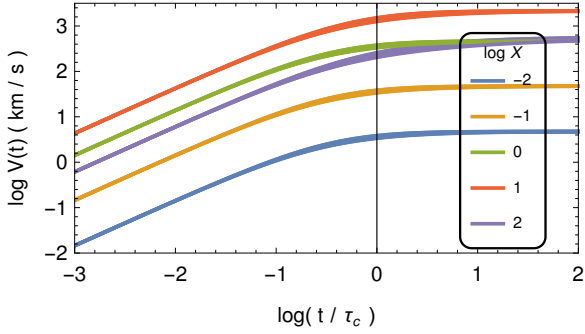}
	\caption{Time evolution of the kick velocity depending on relative quadrupole strength~$X$ for an initial period at birth of $P_0=1$~ms.}
	\label{fig:evolution_vitesse}
\end{figure}

The relation between the rotation rate and time, normalized to the characteristic age $\tau_{\rm c}$ becomes after defining $\mu=\frac{\Omega^2}{\Omega_0^2}$ with $\mu(t=0)=1$
\begin{equation}%\label{key}
\frac{t}{\tau_{\rm c}} =  \frac{1+\xi_0}{2} \, \left[\frac{1}{\mu} - 1 + \xi_0 \, \log \left( \mu \, \frac{1+\xi_0}{1+\mu\,\xi_0}\right)\right] .
\end{equation}
The solution is controlled by the initial spindown ratio $\xi_0$. Fig.~\ref{fig:evolution_vitesse} shows the time evolution of the kick depending on the quadrupole strength depicted by $\xi_0$. The asymptotic velocity is almost reached at the characteristic time $\tau_{\rm c}$. Most efficient kicks are produced near $X\approx10$. The curve shapes are relatively insensitive to the magnitude of~$X$ and~$\xi_0$, but remember that the characteristic time also scales also with $\xi_0$.
While the final kick velocity does not depend on the magnetic field strength because the force acting on the star is proportional to $L$ and the duration of the kick proportional to $1/L$, the characteristic time depends on the field strength through the expression~$\tau_{\rm acc} = \frac{1}{2} \, \frac{I\,\Omega_0^2}{L_0}$
\begin{equation}%\label{key}
\tau_{\rm acc} \approx \numprint{17.4}~\textrm{yr} \, \left( \frac{M}{1.4~M_\odot} \right) \, \left( \frac{P_0}{1~\SIunits{\milli\second}} \right)^2 \, \left( \frac{B}{\numprint{e8}~\SIunits{\tesla}}\right)^{-2} \, \left( \frac{R}{\numprint{12}~\SIunits{\kilo\meter}}\right)^{-4}  .
\end{equation}
This efficient acceleration time scale~$\tau_{\rm acc}$ is shown in Fig.~\ref{fig:temps_acceleration} and compared to the characteristic age for an initial period at birth of $P_0=1$~ms.
This duration on which the electromagnetic kick acts is several orders of magnitude shorter than the characteristic pulsar age~$\tau_{\rm age}$, thus the neutron star had plenty of time to reach the asymptotic value for vanishing angular velocity $a\to0$ or at least when $a\ll a_0$.

If we choose the optimal choice, the acceleration time scale~$\tau_{\rm acc}^{\rm optimal}$ is slightly increased as shown in Fig.~\ref{fig:temps_acceleration}. However, this second scenario can also easily explain the fast proper motion with an electromagnetic kick.

\subsubsection{Young pulsars}

A larger sample of young pulsars is known but only with their transverse proper motion. Neglecting the radial component along the line of sight, we can still obtain good guesses for the real 3D kick velocity by assuming it to be the same order of magnitude as the proper motion. We are aware that the proper motion is also rather uncertain, however it is the best we can compute so far.

\begin{figure}
	\centering
	\includegraphics[width=0.9\linewidth]{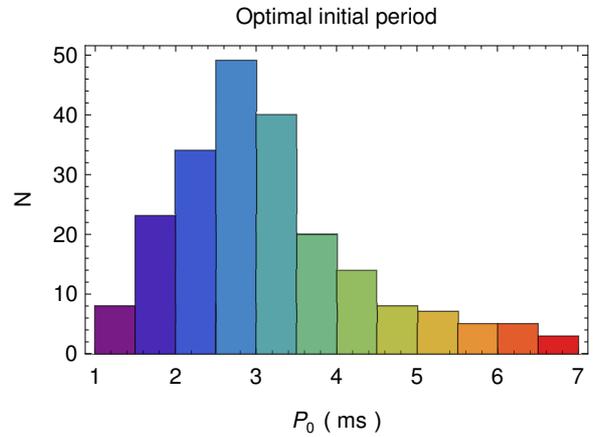}
	\caption{Expected distribution of initial neutron star spin periods~$P_0$ for the sample of young pulsars.}
	\label{fig:P_init_jeune}
\end{figure}

\begin{figure}
	\centering
	\includegraphics[width=0.9\linewidth]{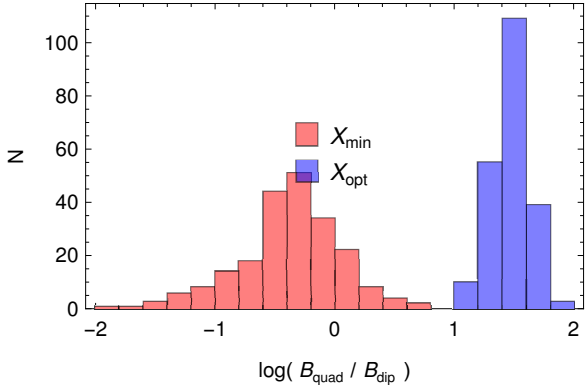}
	\caption{Optimal and minimum quadrupole strength~$X$ distribution of the young pulsars to explain the kick velocity for a period at birth, respectively shown in Fig.~\ref{fig:P_init_jeune} and for $P_0=1$~ms.}
	\label{fig:X_jeune}
\end{figure}

\begin{figure}
	\centering
	\includegraphics[width=0.9\linewidth]{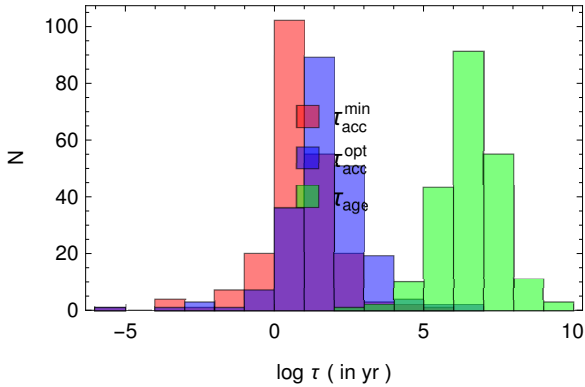}
	\caption{Acceleration time~$\tau_{\rm acc}^{\rm min}$ and~$\tau_{\rm acc}^{\rm opt}$ to the asymptotic kick velocity for young pulsars compared to the characteristic age.}
	\label{fig:temps_acceleration_jeune}
\end{figure}

Following the same procedure as in the previous paragraph, our guess for the initial period of these young pulsars remains in the millisecond range, from 1~ms to 7~ms, as represented in the histogram of Fig.~\ref{fig:P_init_jeune}. The associated quadripolar component strength is shown in blue bars in Fig.~\ref{fig:X_jeune}. It is ten to hundred times larger than the dipole at the surface.
If we assume that all pulsars are born with a 1~ms period, we obtain the graph in red bars of Fig.~\ref{fig:X_jeune}. The quadrupole is now on average two decades weaker than in the previous case.
Finally, the typical acceleration time scales are also shown in red and blue in  Fig.~\ref{fig:temps_acceleration_jeune} and compared to the characteristic age in green. Young pulsars possess plenty of time to move to their current spatial velocities whatever the scenario.

\subsubsection{Millisecond pulsars}

\begin{figure}
	\centering
	\includegraphics[width=0.9\linewidth]{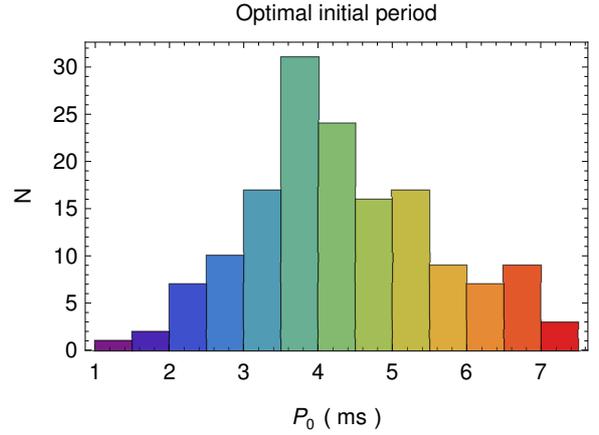}
	\caption{Expected distribution of initial neutron star spin periods~$P_0$ for the sample of millisecond pulsars.}
	\label{fig:P_init_msp}
\end{figure}

\begin{figure}
	\centering
	\includegraphics[width=0.9\linewidth]{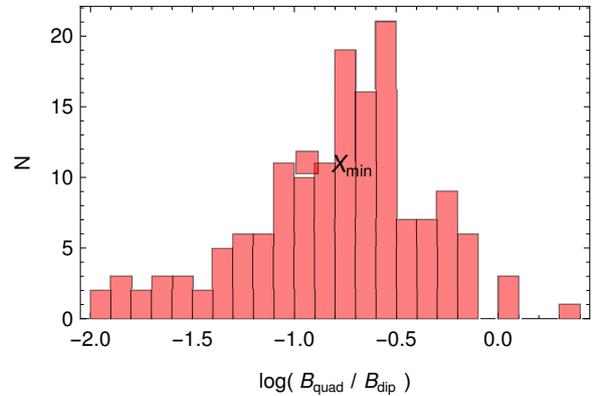}
	\caption{Minimum quadrupole strength~$X$ distribution of the young pulsars to explain the kick velocity for a period at birth shown in Fig.~\ref{fig:P_init_msp}.}
	\label{fig:X_msp}
\end{figure}

\begin{figure}
	\centering
	\includegraphics[width=0.9\linewidth]{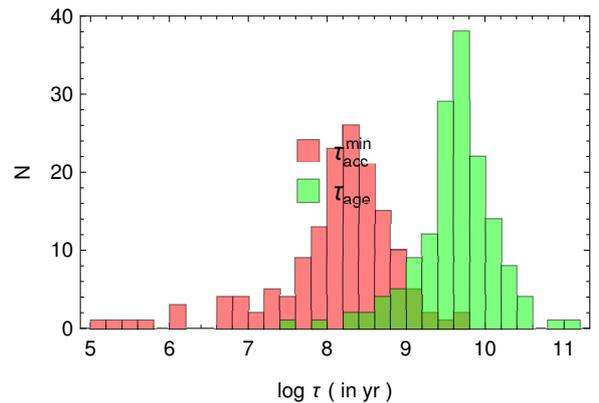}
	\caption{Acceleration time~$\tau_{\rm acc}^{\rm min}$ to the asymptotic kick velocity for millisecond pulsars compared to the characteristic age~$\tau_{\rm age}$.}
	\label{fig:temps_acceleration_msp}
\end{figure}

The same analysis was performed for millisecond pulsars. Results for the most efficient period at beginning of the thrust phase are shown in Fig.~\ref{fig:P_init_msp}. Their correspond to values very close to the current period of these millisecond pulsars. We do not consider initial periods of 1~ms as these pulsars are mostly recycled and spun up during the accretion phase.
The quadrupole strength is plotted in Fig.~\ref{fig:X_msp} and the typical acceleration time compared to the characteristic age in Fig.~\ref{fig:temps_acceleration_msp}. We conclude that here too the current kick velocity was attained at approximately one tenth of their current age.

\section{Conclusions}
\label{sec:Conclusion}

There are more and more compelling evidences that multipolar magnetic fields are anchored in the crust of compact objects like neutron stars thanks to new observations revealing features not expected from a single and/or centred dipole. We computed the exact analytical solutions of a rotating dipole+quadrupole in vacuum with arbitrary relative strengths. Spin down luminosities and electromagnetic forces have been deduced, showing that neutron stars could get significant kick velocities around and above 1000~km/s during their early life if the initial period is less than 10~ms and the quadrupole magnetic field strength comparable to the dipole field or at most one order of magnitude stronger at the surface. This scenario also naturally explains the spin-kick alignment observed in many pulsars. Whereas an off-centred dipole can be ruled out to produce large kicks, a quadridipole generalizing it with an unconstrained quadrupole offers an attractive alternative. We showed that the birth period of neutron stars sensitively depend on the magnetic braking law. Time-dependent parameters in the spin down luminosity can falsify the traditional estimate found from the neutron star/supernova association.

Neutron stars are however usually surrounded by a relativistic plasma screening a significant part of the electric field, influencing the spin down and electromagnetic torque. We plan to extend the current vacuum model to force-free quadridipole fields which seem important to model thermal X-ray emission from the surface as measured by NICER \citep{bilous_nicer_2019}.

\section*{Acknowledgements}

I am grateful to the referee for helpful comments and suggestions. This work has been supported by the CEFIPRA grant IFC/F5904-B/2018.

The data underlying this article will be shared on reasonable request to the corresponding author.

%%%%%%%%%%%%%%%%%%%%%%%%%%%%%%%%%%%%%%%%%%%%%%%%%%



%\bibliographystyle{mn2e}
%\bibliography{/home/petri/zotero/Ma_bibliotheque}

\begin{thebibliography}{35}
\expandafter\ifx\csname natexlab\endcsname\relax\def\natexlab#1{#1}\fi

\bibitem[{AlGendy \& Morsink(2014)}]{algendy_universality_2014}
AlGendy M., Morsink S.~M., 2014, ApJ, 791, 78

\bibitem[{Arfken \& Weber(2005)}]{arfken_mathematical_2005}
Arfken G.~B., Weber H.-J., 2005, Mathematical methods for physicists, 6th edn.
  Elsevier, Boston

\bibitem[{Atoyan(1999)}]{atoyan_radio_1999}
Atoyan A.~M., 1999, A\&A, 346, L49

\bibitem[{Belvedere {et~al}\mbox{.}(2014)Belvedere, Boshkayev, Rueda, \&
  Ruffini}]{belvedere_uniformly_2014}
Belvedere R., Boshkayev K., Rueda J.~A., Ruffini R., 2014, Nuclear Physics A,
  921, 33

\bibitem[{Bilous {et~al}\mbox{.}(2019)Bilous, Watts, Harding, Riley,
  Arzoumanian, Bogdanov, Gendreau, Ray, Guillot, Ho, \&
  Chakrabarty}]{bilous_nicer_2019}
Bilous A.~V. {et~al.}, 2019, ApJL, 887, L23

\bibitem[{Chandrasekhar(1970)}]{chandrasekhar_ellipsoidal_1970}
Chandrasekhar S., 1970, Ellipsoidal {Figures} of {Equilibrium}. Yale University
  Press, New Haven

\bibitem[{Chen {et~al}\mbox{.}(2020)Chen, Yuan, \&
  Vasilopoulos}]{chen_numerical_2020}
Chen A.~Y., Yuan Y., Vasilopoulos G., 2020, ApJL, 893, L38

\bibitem[{Cordes \& Chernoff(1998)}]{cordes_neutron_1998}
Cordes J.~M., Chernoff D.~F., 1998, ApJ, 505, 315

\bibitem[{Ferrario \& Wickramasinghe(2007)}]{ferrario_birth_2007}
Ferrario L., Wickramasinghe D., 2007, MNRAS, 375, 1009

\bibitem[{Finn \& Shapiro(1990)}]{finn_spin-down_1990}
Finn L.~S., Shapiro S.~L., 1990, ApJ, 359, 444

\bibitem[{Friedman {et~al}\mbox{.}(1989)Friedman, Ipser, \&
  Parker}]{friedman_implications_1989}
Friedman J.~L., Ipser J.~R., Parker L., 1989, Phys. Rev. Lett., 62, 3015

\bibitem[{Gralla {et~al}\mbox{.}(2017)Gralla, Lupsasca, \&
  Philippov}]{gralla_inclined_2017}
Gralla S.~E., Lupsasca A., Philippov A., 2017, ApJ, 851, 137

\bibitem[{Haensel {et~al}\mbox{.}(1995)Haensel, Salgado, \&
  Bonazzola}]{haensel_equation_1995}
Haensel P., Salgado M., Bonazzola S., 1995, A\&A, 296,
  745

\bibitem[{Harrison \& Tademaru(1975)}]{harrison_acceleration_1975}
Harrison E.~R., Tademaru E., 1975, ApJ, 201, 447

\bibitem[{Heger {et~al}\mbox{.}(2000)Heger, Langer, \&
  Woosley}]{heger_presupernova_2000}
Heger A., Langer N., Woosley S.~E., 2000, ApJ, 528, 368

\bibitem[{Heger {et~al}\mbox{.}(2005)Heger, Woosley, \&
  Spruit}]{heger_presupernova_2005}
Heger A., Woosley S.~E., Spruit H.~C., 2005, ApJ, 626, 350

\bibitem[{Johnston {et~al}\mbox{.}(2005)Johnston, Hobbs, Vigeland, Kramer,
  Weisberg, \& Lyne}]{johnston_evidence_2005}
Johnston S., Hobbs G., Vigeland S., Kramer M., Weisberg J.~M., Lyne A.~G.,
  2005, MNRAS, 364, 1397

\bibitem[{Johnston {et~al}\mbox{.}(2007)Johnston, Kramer, Karastergiou, Hobbs,
  Ord, \& Wallman}]{johnston_evidence_2007}
Johnston S., Kramer M., Karastergiou A., Hobbs G., Ord S., Wallman J., 2007, MNRAS, 381, 1625

\bibitem[{Kojima \& Kato(2011)}]{kojima_kick_2011}
Kojima Y., Kato Y.~E., 2011, ApJ, 728, 75

\bibitem[{Lai {et~al}\mbox{.}(2001)Lai, Chernoff, \& Cordes}]{lai_pulsar_2001}
Lai D., Chernoff D.~F., Cordes J.~M., 2001, ApJ, 549,
  1111

\bibitem[{Manchester {et~al}\mbox{.}(2005)Manchester, Hobbs, Teoh, \&
  Hobbs}]{manchester_australia_2005}
Manchester R.~N., Hobbs G.~B., Teoh A., Hobbs M., 2005, The Astronomical
  Journal, 129, 1993

\bibitem[{Manchester \& Taylor(1977)}]{manchester_pulsars_1977}
Manchester R.~N., Taylor J.~H., 1977, San Francisco : W. H. Freeman, c1977.

\bibitem[{Marshall {et~al}\mbox{.}(1998)Marshall, Gotthelf, Zhang, Middleditch,
  \& Wang}]{marshall_discovery_1998}
Marshall F.~E., Gotthelf E.~V., Zhang W., Middleditch J., Wang Q.~D., 1998, ApJL, 499, L179

\bibitem[{Michel \& Goldwire(1970)}]{michel_alignment_1970}
Michel F.~C., Goldwire, Jr. H.~C., 1970, Astrophysical Letters, 5, 21

\bibitem[{Morsink {et~al}\mbox{.}(2007)Morsink, Leahy, Cadeau, \&
  Braga}]{morsink_oblate_2007}
Morsink S.~M., Leahy D.~A., Cadeau C., Braga J., 2007, ApJ, 663, 1244

\bibitem[{Ott {et~al}\mbox{.}(2006)Ott, Burrows, Thompson, Livne, \&
  Walder}]{ott_spin_2006}
Ott C.~D., Burrows A., Thompson T.~A., Livne E., Walder R., 2006, ApJS, 164,
  130

\bibitem[{P{\'e}tri(2013)}]{petri_general-relativistic_2013}
P{\'e}tri J., 2013, MNRAS, 433, 986

\bibitem[{P{\'e}tri(2015)}]{petri_multipolar_2015}
P{\'e}tri J., 2015, MNRAS, 450, 714

\bibitem[{P{\'e}tri(2016)}]{petri_radiation_2016}
P{\'e}tri J., 2016, MNRAS, 463, 1240

\bibitem[{P{\'e}tri(2019)}]{petri_illusion_2019}
P{\'e}tri J., 2019, MNRAS, 485, 4573

\bibitem[{P{\'e}tri \& Mitra(2020)}]{petri_joint_2020}
P{\'e}tri J., Mitra D., 2020, MNRAS, 491, 80

\bibitem[{Putney \& Jordan(1995)}]{putney_off-centered_1995}
Putney A., Jordan S., 1995, ApJ, 449, 863

\bibitem[{Rankin(2007)}]{rankin_further_2007}
Rankin J.~M., 2007, ApJ, 664, 443

\bibitem[{Riley {et~al}\mbox{.}(2019)Riley, Watts, Bogdanov, Ray, Ludlam,
  Guillot, Arzoumanian, Baker, Bilous, Chakrabarty, Gendreau, Harding, Ho,
  Lattimer, Morsink, \& Strohmayer}]{riley_nicer_2019}
Riley T.~E. {et~al.}, 2019, ApJL, 887, L21

\bibitem[{Roberts(1979)}]{roberts_electromagnetic_1979}
Roberts W.~J., 1979, ApJS, 41, 75

\end{thebibliography}


%%%%%%%%%%%%%%%%%%%%%%%%%%%%%%%%%%%%%%%%%%%%%%%%%%

%%%%%%%%%%%%%%%%% APPENDICES %%%%%%%%%%%%%%%%%%%%%

% \appendix
% 
% \section{Some extra material}

%%%%%%%%%%%%%%%%%%%%%%%%%%%%%%%%%%%%%%%%%%%%%%%%%%

% Don't change these lines
\bsp	% typesetting comment
\label{lastpage}
\end{document}